\documentclass[aps,pre,reprint]{revtex4-2}
\usepackage{bm,cases}
\usepackage{graphicx,color}

\begin{document}

\title{A minimal model with stochastically broken reciprocity}

\author{Z. C. Tu}
\email{tuzc@bnu.edu.cn}
\affiliation{School of Physics and Astronomy, Beijing Normal University, Beijing 100875, China}
\affiliation{Key Laboratory of Multiscale Spin Physics (Ministry of
Education) Beijing Normal University, Beijing 100875}

\date{\today}
\begin{abstract}
We introduce a minimal model consisting of a two-body system with stochastically broken reciprocity (i.e., random violation of Newton's third law) and then investigate its statistical behaviors, including fluctuations of velocity and position, time evolution of probability distribution functions, energy gain, and entropy production. The effective temperature of this two-body system immersed in a thermal bath is also derived. Furthermore, we heuristically present an extremely minimal model where the relative motion adheres to the same rules as in classical mechanics, while the effect of stochastically broken reciprocity only manifests in the fluctuating motion of the center of mass.
\end{abstract}

\maketitle

\section{Introduction}
Newton's laws of motion are the cornerstone of classical mechanics. Among them, the first law defines inertial frames; the second law establishes the relationship between acceleration and force; and the third law describes the reciprocity of action and reaction forces. Here, reciprocity means that the action force exerted by particle A on particle B and the reaction force exerted by B on A are equal in magnitude, opposite in direction, and collinear with the straight line connecting A and B.
This reciprocal nature between action and reaction forces governs not only fundamental microscopic interactions but also the emergent forces between passive particles in equilibrium media~\cite{Israelachvili1992,Likos2001}. However, reciprocity is found to be broken in non-equilibrium systems where the emergent action and reaction forces are either unequal in magnitude, or out of collinearity~\cite{Ivlev2015,PoncetPRX048002}. Typically broken phenomena manifest in active systems~\cite{Vicsek1995,MarchettiRMP2013,BechingerRMP2016,BowickPRX2022,ChenPRE96020601,Dadhichi101052601,KreienkampNJP2022}, especially in micro-swimmers~\cite{YasudaPRE062113,YasudaJPSJ075001,Golestanian062901,Golestanian036308}, active colloids~\cite{SpeckPRL2017,SchmidtJCP2019,MeredithNC2020,WuNN2021-288,GardiPRL2023}, and robotic systems~\cite{Lavergne2019,ZhangHPPRL2024,LiuCPB014101,WangPRL2021108002}.

The strict breaking of reciprocity leads to odd viscosity or elasticity~\cite{Banerjee2017,Scheibner2020,FruchartARCMP2023,HosakaPRE2021,ZhaoFP2022,LinPRR2024,KobayashiJPSJ2023}, unconventional phase transitions~\cite{TonerTu1995,Fruchart2021,TanNat2022,WangYTCTP067601,ChenTu115603,Chenyong075601,Liyycpl2023}, and exotic transport behaviors~\cite{YangMC198001,ShiXQPRL258302,Zhengzg2025APS,AiBQPRE064409}.
Instead of delving into the aforementioned in-depth discussions on the consequences of strictly broken reciprocity in active or nonequilibrium systems, we adopt an inverted perspective to explore how insights gleaned from the study of nonequilibrium thermodynamics and active matter can illuminate our rethinking of the fundamental laws of classical mechanics. It is widely accepted that fundamental forces in classical mechanics strictly abide by Newton's third law. Experimental observations by scientists typically focus on the relationships between the mean values of physical quantities. Theoretically, however, we cannot entirely rule out the possibility that certain fundamental forces might, on average, satisfy Newton's third law but stochastically violate it if the vacuum--regarded as an active ``aether"--is not in equilibrium. Introducing such stochastic violations into fundamental interactions could have profound implications across fields from quantum mechanics and particle physics to cosmology, though this remains a speculative direction requiring rigorous theoretical and experimental scrutiny.
Since the third law holds on average, we anticipate that the core conclusions of classical mechanics remain valid when expressed in terms of the mean values of relevant physical quantities. Stochastically broken reciprocity, however, manifests its effects in the fluctuations of these physical quantities. Driven by theoretical interest and pure curiosity, we pose the question: How does classical mechanics retain its validity at the level of mean values, while stochastic violations leave their imprint on fluctuations?

To the best of the author's knowledge, the concept of stochastically broken reciprocity--defined as the random violation of Newton's third law--has received scant attention in both historical and recent studies, with the sole exception of the insightful work by Cocconi, Alston, and Bertrand~\cite{CocconiPRS2023}. Their research introduces linear pairwise interactions that exhibit reciprocity-breaking fluctuations around a reciprocal mean coupling strength. However, since the interaction term is defined as the product of coupling strength and coordinates, it remains a subtle issue whether Newton's third law holds on average within the Cocconi-Alston-Bertrand model. This ambiguity makes a request for a more precise and minimal model that incorporates stochastically broken reciprocity while ensuring that violations of Newton's third law are minimized. In this paper, we aim to fill this gap by introducing a minimal model of a two-body system with stochastically broken reciprocity, while ensuring that Newton's third law holds on average. We then systematically investigate its statistical signatures such as fluctuations of velocity and position, the evolution of probability distribution functions (PDFs), energy gain, entropy production, and so on. The remainder of this paper is organized as follows. In Sec.~\ref{sec-minmodel}, we specifically describe a two-particle system with stochastically broken reciprocity, and derive the dynamic equations of motion for the system's center of mass (COM) and the two-body relative position based on Newton's second law. In Sec.~\ref{sec-correfunc}, we compute the mean square velocity and mean square displacement of the COM, and the covariance matrix of the two-body relative motion with a deterministic harmonic interaction. In Sec.~\ref{sec-evolpdf}, we derive three Fokker-Planck equations governing the PDFs of the COM motion, the relative motion, and their joint evolution, and use them to track the changes of energy and entropy. In Sec.~\ref{sec-inbaths}, we immerse the system in a thermal bath and, under overdamped conditions, obtain the Smoluchowski equation governing the PDF of the relative position. In Sec.~\ref{sec-mostminmodel}, we present an extremely minimal model  in which the relative motion obeys classical mechanics exactly, while the stochastically broken reciprocity merely influence the fluctuating motion of the COM. The final section provides a brief summary and outlook.

\section{minimal model\label{sec-minmodel}}
\begin{figure}[!htp]
\includegraphics[width=4cm]{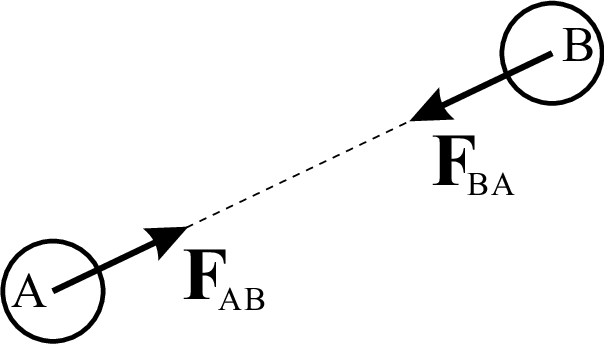}
\caption{Two-body system. Two particles are labeled with A and B, respectively. $\mathbf{F}_\mathrm{BA}$ represents the action force exerted by A on B, while $\mathbf{F}_\mathrm{AB}$ represents the reaction force exerted by B on A.\label{fig-twobody}}
\end{figure}

Consider a two-body system as depicted in Fig.~\ref{fig-twobody}. Particles A and B, with masses $m_\mathrm{A}$ and $m_\mathrm{B}$, have instantaneous positions $\mathbf{r}_\mathrm{A}$ and $\mathbf{r}_\mathrm{B}$ measured in an inertial frame. The force exerted by A on B is $\mathbf{F}_\mathrm{BA}$; the corresponding reaction force exerted by B on A is $\mathbf{F}_\mathrm{AB}$.

We assume that Newton's third law holds on average. Mathematically, this is expressed as
\begin{equation}\label{eq-avg-recpro}
\left\langle\mathbf{F}_\mathrm{AB}\right\rangle=-\left\langle\mathbf{F}_\mathrm{BA}\right\rangle,~\mathrm{and}~\left\langle\mathbf{F}_\mathrm{BA}\right\rangle \parallel \overline{AB}
\end{equation}
where $\overline{AB}$ denotes the line connecting particles A and B. To render the statement unambiguous, we decompose the instantaneous forces as
\begin{numcases}{}
 \mathbf{F}_\mathrm{BA} =-\nabla \phi(r)+\sqrt{g_\mathrm{B}}\bm{\xi}_\mathrm{B}(t),\label{eq-FBA}\\
 \mathbf{F}_\mathrm{AB} =\nabla \phi(r)+\sqrt{g_\mathrm{A}}\bm{\xi}_\mathrm{A}(t),\label{eq-FAB}
\end{numcases}
where $r$ is the distance between particles A and B, and $\phi$ is a scalar function of $r$.
The constants $\sqrt{g_\mathrm{A}}$ and $\sqrt{g_\mathrm{B}}$ (with dimension of momentum, i.e., Mass$\cdot$Length$\cdot$Time$^{-1}$) quantify the magnitude of reciprocity violation. We argue that $g_\mathrm{A}$ and $g_\mathrm{B}$ simultaneously depend on certain unknown properties of both A and B, rather than on the property of just one of them alone. The Gaussian white noises $\bm{\xi}_\alpha(t)$ ($\alpha=\mathrm{A,B}$) satisfy
\begin{equation}\label{eq-noisemean}
\left\langle\bm{\xi}_\alpha (t)\right\rangle=0,
\end{equation}
and
\begin{equation}\label{eq-noisecorrelate}
\left\langle\bm{\xi}_\alpha (t)\bm{\xi}^\mathrm{T}_\beta(t')\right\rangle=\delta_{\alpha \beta}\delta(t-t')\mathbf{I},~(\alpha,\beta=\mathrm{A,B})
\end{equation}
with $\delta_{\alpha \beta}$ the Kronecker symbol, $\delta(t-t')$ the Dirac delta function, $\mathbf{I}$ the unit tensor. The superscript ``$\mathrm{T}$" represents the transpose operation on vectors or matrices.

It is evident that Eqs.~(\ref{eq-FBA}) and (\ref{eq-FAB}) give $\mathbf{F}_\mathrm{BA}+\mathbf{F}_\mathrm{AB}=\sqrt{g_\mathrm{B}}\bm{\xi}_\mathrm{B}(t)+\sqrt{g_\mathrm{A}}\bm{\xi}_\mathrm{A}(t)$, which generally does not vanish. As a consequence, Newton's third law is violated due to the stochastic term $\sqrt{g_\mathrm{B}}\bm{\xi}_\mathrm{B}(t)+\sqrt{g_\mathrm{A}}\bm{\xi}_\mathrm{A}(t)$. However, Eq.~(\ref{eq-noisemean}) ensures the satisfaction of the average relation~(\ref{eq-avg-recpro}), thereby preserving Newton's third law in the statistical sense. The dynamical system governed by equations~(\ref{eq-FBA})--(\ref{eq-noisecorrelate}) thus represents a minimal stochastic violation of Newton's third law. In this context, we refer to this framework a minimal model with stochastically broken reciprocity.

Naturally, there exists an alternative scheme for implementing stochastically broken reciprocity. As an illustration, consider replacing Eqs.~(\ref{eq-FBA}) and (\ref{eq-FAB}) with
\begin{numcases}{}
 \mathbf{F}_\mathrm{BA} =-\nabla \phi(r)[1+\sqrt{g_\mathrm{B}}{\xi}_\mathrm{B}(t)],\label{eq-revise-FBA}\\
 \mathbf{F}_\mathrm{AB} =\nabla \phi(r)[1+\sqrt{g_\mathrm{A}}{\xi}_\mathrm{A}(t)].\label{eq-revise-FAB}
\end{numcases}
The Cocconi-Alston-Bertrand model~\cite{CocconiPRS2023} provides a concrete one-dimensional example where the aforementioned potential $\phi$ takes a harmonic form. Importantly, one cannot simply claim that $\langle \nabla \phi(r) {\xi}_\mathrm{A}(t)\rangle=0$ and $\langle \nabla \phi(r) {\xi}_\mathrm{B}(t)\rangle=0$, as illustrated by the counterexample in Appendix~\ref{sec-appendA}. Consequently, this scheme cannot definitely guarantee that Newton's third law holds on average. In what follows, we will focus on the simpler framework that preserves the average validity of Newton's third law, with subsequent analyses based on the minimal model defined by Eqs.~(\ref{eq-FBA})--(\ref{eq-noisecorrelate}).

According to Newton's second law, the equations of motion for the two particles are
\begin{numcases}{}
 m_\mathrm{B}\ddot{\mathbf{r}}_\mathrm{B} =-\nabla \phi(r)+\sqrt{g_\mathrm{B}}\bm{\xi}_\mathrm{B}(t),\label{eq-motionB}\\
 m_\mathrm{A}\ddot{\mathbf{r}}_\mathrm{A} =\nabla \phi(r)+\sqrt{g_\mathrm{A}}\bm{\xi}_\mathrm{A}(t),\label{eq-motionA}
\end{numcases}

Introduce the COM coordinate
\begin{equation}\label{eq-masscenter}
\mathbf{R}\equiv\frac{m_\mathrm{A} \mathbf{r}_\mathrm{A}+m_\mathrm{B} \mathbf{r}_\mathrm{B}}{M}
\end{equation}
with $M\equiv m_\mathrm{A}+m_\mathrm{B}$ being the total mass of the system.
Define the relative position vector of B with respect to A
\begin{equation}\label{eq-relatvector}
\mathbf{r}\equiv\mathbf{r}_\mathrm{B}- \mathbf{r}_\mathrm{A}.
\end{equation}
Then Eqs.~(\ref{eq-motionB}) and (\ref{eq-motionA}) yield stochastic differential equations
\begin{numcases}{}
\ddot{\mathbf{R}}= \bm{\zeta}_1(t), \label{eq-motCM}\\
\ddot{\mathbf{r}}= -\frac{\nabla \phi}{\mu} + \bm{\zeta}_2(t), \label{eq-motrelative}
\end{numcases}
with $\mu\equiv m_\mathrm{A}m_\mathrm{B}/M$ being the reduced mass of the system. The effective noises are
\begin{numcases}{}
\bm{\zeta}_1(t)= \frac{\sqrt{g_\mathrm{A}}}{M}\bm{\xi}_\mathrm{A}(t)+\frac{\sqrt{g_\mathrm{B}}}{M}\bm{\xi}_\mathrm{B}(t), \label{eq-zeta1}\\
\bm{\zeta}_2(t)= - \frac{\sqrt{g_\mathrm{A}}}{m_\mathrm{A}}\bm{\xi}_\mathrm{A}(t)+\frac{\sqrt{g_\mathrm{B}}}{m_\mathrm{B}}\bm{\xi}_\mathrm{B}(t). \label{eq-zeta2}
\end{numcases}
From Eq.~(\ref{eq-noisemean}) we find that these effective noises have vanishing mean:
\begin{equation}\label{eq-noisezetamean}
\left\langle\bm{\zeta}_i (t)\right\rangle=0,~(i=1,2).
\end{equation}
In addition, from Eq.~(\ref{eq-noisecorrelate}) we obtain their correlations
\begin{equation}\label{eq-zetacorrelate}
\left\langle\bm{\zeta}_i (t)\bm{\zeta}^\mathrm{T}_j(t')\right\rangle=g_{ij}\delta(t-t')\mathbf{I},~(i,j=1,2)
\end{equation}
where the coefficients are explicitly expressed as
\begin{numcases}{}
  g_{11} = \frac{g_\mathrm{A}+g_\mathrm{B}}{M^2}, \label{eq-g11}\\
  g_{12} = g_{21}= \frac{1}{M}(\frac{g_\mathrm{B}}{m_\mathrm{B}}-\frac{g_\mathrm{A}}{m_\mathrm{A}}),  \label{eq-g12}\\
  g_{22} = \frac{g_\mathrm{A}}{m_\mathrm{A}^2}+\frac{g_\mathrm{B}}{m_\mathrm{B}^2}.  \label{eq-g22}
\end{numcases}

Stochastic differential equations (\ref{eq-motCM}) and (\ref{eq-motrelative}) combining with (\ref{eq-noisezetamean}) and (\ref{eq-zetacorrelate}) govern the COM motion and the relative motion of B with respect A.

\section{Fluctuations of velocity and position\label{sec-correfunc}}
The observable consequences of stochastically broken reciprocity are encoded in the fluctuation behaviors of the system. In what follows we examine the mean square velocity and displacement for the COM motion, and the covariance matrix for the relative motion of the two particles. To be concrete, we restrict the discussion to one spatial dimension and take the interaction potential to be of harmonic form.

\subsection{Mean square velocity and displacement for the COM motion}
In one-dimensional situation, let $X$ and $V$ denote the position and velocity of the COM. From Eq.~(\ref{eq-motCM}) together with Eqs.~(\ref{eq-noisezetamean}) and (\ref{eq-zetacorrelate}), we obtain the equations governing the COM motion
\begin{numcases}{}
  \dot{X} = V, \label{eq-motCM1DX}\\
  \dot{V} = \zeta_1(t), \label{eq-motCM1DV}
\end{numcases}
where the Gaussian white noise ${\zeta}_1 (t)$ satisfies
\begin{equation}\label{eq-noisezetamean1D1}
\left\langle{\zeta}_1 (t)\right\rangle=0,
\end{equation}
and
\begin{equation}\label{eq-zetacorrelate1D1}
\left\langle{\zeta}_1 (t){\zeta}_1(t')\right\rangle=g_{11}\delta(t-t').
\end{equation}

By integrating Eq.~(\ref{eq-motCM1DV}), we immediately obtain the COM velocity
\begin{equation}\label{eq-solmotCM1DV}
 V = V_0+\int_0^t\zeta_1(\tau)\mathrm{d}\tau,
\end{equation}
where $V_0$ represents the initial velocity of the COM. By integrating the above equation again and exchanging the order of integration for double integral, we can obtain the COM displacement
\begin{equation}
  \Delta X\equiv X-X_0 = V_0t+\int_0^t(t-\tau)\zeta_1(\tau)\mathrm{d}\tau, \label{eq-solmotCM1DX}
\end{equation}
with $X_0$ the initial position of the COM. Considering Eq.~(\ref{eq-noisezetamean1D1}), we immediately obtain $\left\langle V\right\rangle=\left\langle V_{0}\right\rangle$ and $\left\langle \Delta X \right\rangle =\left\langle V_{0}\right\rangle t$ from Eqs.~(\ref{eq-solmotCM1DV}) and (\ref{eq-solmotCM1DX}). Hence, on average, the COM executes uniform rectilinear motion.

We now calculate the mean square velocity. Squaring Eq.~(\ref{eq-solmotCM1DV}) and considering Eq.~(\ref{eq-noisezetamean1D1}), we have
\begin{equation}\label{eq-Volet2}
\langle V^2\rangle=\langle V_0^2 \rangle +\left\langle\left[\int_0^t\zeta_1(\tau)\mathrm{d}\tau\right]^2\right\rangle.
\end{equation}
To deal with the second term on the right-handed side of above equation, we transform the square of the integral $[\int_0^t\zeta_1(\tau)d\tau]^2$ into a double integral $\int_0^t\int_0^t \zeta_1(\tau)\zeta_1(\tau')\mathrm{d}\tau\mathrm{d}\tau'$. Invoking Eq.~(\ref{eq-zetacorrelate1D1}), we further derive $\langle[\int_0^t\zeta_1(\tau)\mathrm{d}\tau]^2\rangle=\int_0^t\int_0^t \langle\zeta_1(\tau)\zeta_1(\tau')\rangle \mathrm{d}\tau\mathrm{d}\tau'=\int_0^t\int_0^t g_{11}\delta(\tau-\tau')\mathrm{d}\tau\mathrm{d}\tau'=\int_0^t g_{11} \mathrm{d}\tau=g_{11} t$. Substituting this result into Eq.~(\ref{eq-Volet2}), we finally obtain the mean square velocity:
\begin{equation}\label{eq-corlateV2}
\langle V^2\rangle=\langle V_0^2 \rangle+g_{11}t.
\end{equation}
From this equation, we can directly write out the fluctuation of velocity, $\langle (V-\langle V\rangle)^2\rangle=\langle (V_0-\langle V_0\rangle)^2\rangle +g_{11}t$, which implies that the fluctuation of velocity increases linearly with time. The linear time-dependent term involving $g_{11}$ in Eq.~(\ref{eq-corlateV2}) embodies the influence of stochastically broken reciprocity. When $g_{11}=0$, this result simplifies to $\langle V^2\rangle=\langle V_0^2 \rangle$, corresponding to the familiar scenario of uniform rectilinear motion in classical mechanics.

\begin{figure}[!htp]
\includegraphics[width=7cm]{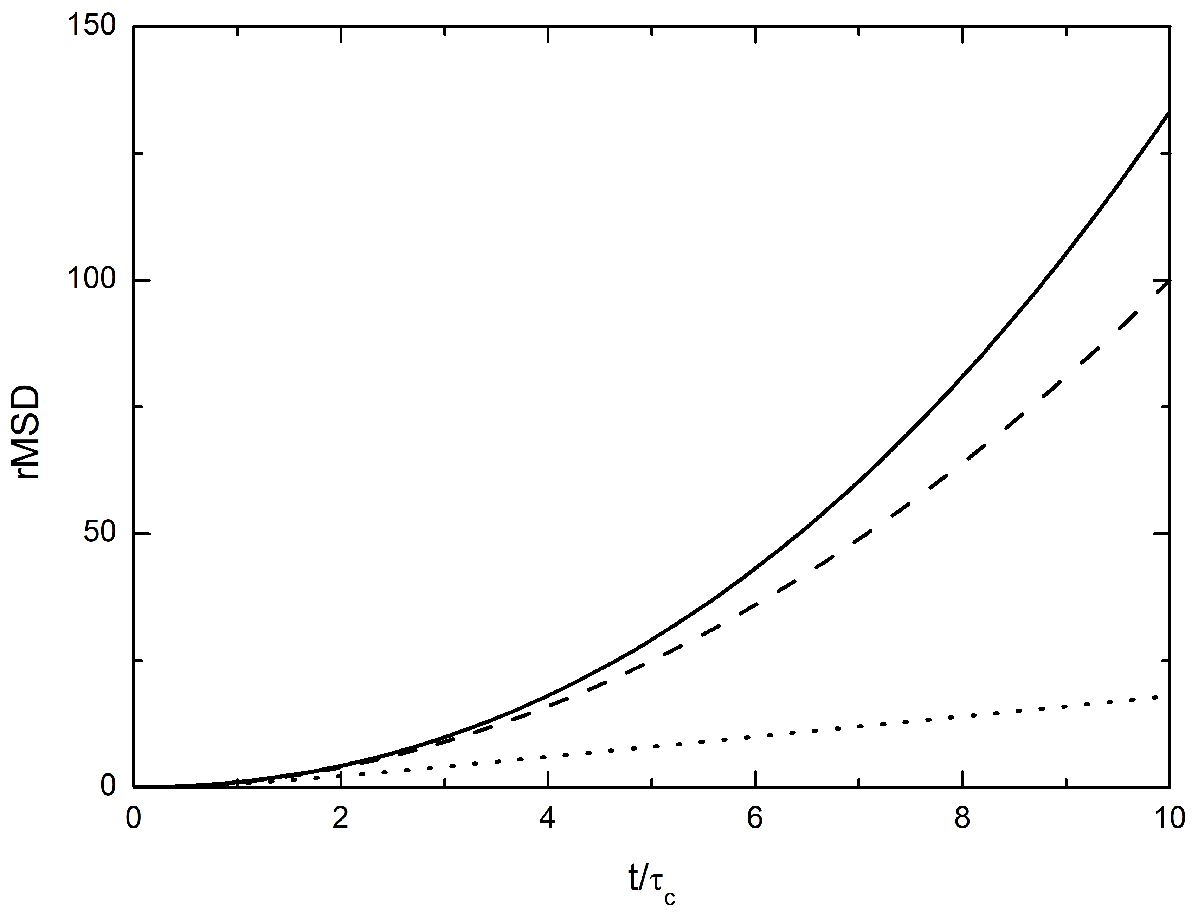}
\caption{Time dependence of reduced mean square displacements (rMSD). The solid, dashed and dotted lines correspond to $\left\langle (\Delta X)^2\right\rangle/D\tau_c$, $\left\langle (\Delta X)^2\right\rangle_\mathrm{CM}/D\tau_c$, and $\left\langle (\Delta X)^2\right\rangle_\mathrm{BM}/D\tau_c$, respectively. For the plotting of these curves, the parameters employed are $\left\langle V_0^2\right\rangle=D/\tau_c$ and $g_{11}=D/10\tau_c^2$.\label{fig2-msd}}
\end{figure}

Repeating the similar procedure, we can achieve the mean square displacement:
\begin{equation}\label{eq-corlateXX}
\left\langle (\Delta X)^2\right\rangle =\left\langle V^2_{0}\right\rangle t^2+\frac{g_{11}}{3}t^3.
\end{equation}
From the above equation, we can directly write out the fluctuation of displacement, $\langle (\Delta X-\langle \Delta X\rangle)^2\rangle=\langle (V_0-\langle V_0\rangle)^2\rangle t^2 +({g_{11}}/{3})t^3$. The cubic time-dependent term containing $g_{11}$ in Eq.~(\ref{eq-corlateXX}) reflects the influence of stochastically broken reciprocity. When $g_{11}=0$, this result reduces to $\left\langle (\Delta X)^2\right\rangle_\mathrm{CM} =\left\langle V^2_{0}\right\rangle t^2$, corresponding to the well-known scenario of uniform rectilinear motion in classical mechanics. Furthermore, the cubic law is distinctly different from the Brownian motion in a thermal bath, where the mean square displacement can be expressed as:
\begin{equation}\label{eq-Brown-mot}
\left\langle (\Delta X)^2\right\rangle_\mathrm{BM} =2 D [t-\tau_c(1-\mathrm{e}^{-t/\tau_c})],
\end{equation}
where $D$ and $\tau_c$ denote diffusion coefficient and momentum relaxation time, respectively~\cite{Reichlbook}. It is straightforward to observe that $\left\langle (\Delta X)^2\right\rangle_\mathrm{BM}\sim t^2$ in the short-time regime, whereas $\left\langle (\Delta X)^2\right\rangle_\mathrm{BM}\sim t$ in the long-time limit. The time dependence of three mean square displacements $\left\langle (\Delta X)^2\right\rangle$, $\left\langle (\Delta X)^2\right\rangle_\mathrm{CM}$ and $\left\langle (\Delta X)^2\right\rangle_\mathrm{BM}$ is schematically illustrated in Fig.~\ref{fig2-msd}. A key observation is that the pronounced discrepancies in their long-time behaviors offer an experimental criterion for distinguishing systems with stochastically broken reciprocity from two reference cases: classical systems that satisfy Newton's third law, and Brownian particles immersed in thermal baths.

\subsection{Covariance matrix for the relative motion of B with respect to A}
In one-dimensional situation, the relative position and velocity of B with respect to A are denoted as $x$ and $v$, respectively. According to Eqs.~(\ref{eq-motrelative}), (\ref{eq-noisezetamean}) and (\ref{eq-zetacorrelate}), the equations of relative motion of B with respect to A can be expressed as
\begin{numcases}{}
  \dot{x} = v, \label{eq-relatposi}\\
  \dot{v} =-\phi'/\mu + \zeta_2(t), \label{eq-relatvelc}
\end{numcases}
where $\phi'\equiv \mathrm{d}\phi/\mathrm{d}x$, and $\zeta_2 (t)$ is a Gaussian white noise satisfying
\begin{equation}\label{eq-noisezetamean1D2}
\left\langle{\zeta}_2 (t)\right\rangle=0,
\end{equation}
and
\begin{equation}\label{eq-zetacorrelate1D2}
\left\langle{\zeta}_2 (t){\zeta}_2(t')\right\rangle=g_{22}\delta(t-t').
\end{equation}

For simplicity, we take the harmonic potential $\phi=\frac{1}{2}\mu \omega^2 x^2$ so that Eq.~(\ref{eq-relatvelc}) reduces to
\begin{equation}
\dot{v} =-\omega^2 x + \zeta_2(t). \label{eq-relatveloscl}
\end{equation}

Equations~(\ref{eq-relatposi}) and (\ref{eq-relatveloscl}) describe a linear, damp-free harmonic oscillator driven by the stochastic force $\zeta_2(t)$. Following the standard procedure, we achieve the solution:
\begin{equation}\label{eq-oscilator}
\left(
  \begin{array}{c}
     v\\
    \omega x \\
  \end{array}
\right)= U \left(
  \begin{array}{c}
     v_0\\
    \omega x_0 \\
  \end{array}
\right) + \int_0^t\left(
  \begin{array}{c}
\cos\omega(t-\tau) \\
   \sin\omega(t-\tau) \\
  \end{array}
\right)\zeta_2(\tau)\mathrm{d}\tau
\end{equation}
with the rotation matrix $U=\left(
  \begin{array}{cc}
    \cos\omega t & -\sin \omega t\\
    \sin \omega t & \cos\omega t \\
  \end{array}
\right)$. Here $v_0$ and $x_0$ denote respectively the initial relative velocity and position of B with respect to A.
Taking the average on Eq.~(\ref{eq-oscilator}) and using Eq.~(\ref{eq-noisezetamean1D2}), we immediately obtain \begin{equation}\label{eq-oscilatormean}
\left(
  \begin{array}{c}
    \langle v\rangle\\
    \omega\langle x\rangle \\
  \end{array}
\right)=U \left(
  \begin{array}{c}
    \langle v_0\rangle\\
    \omega\langle x_0\rangle \\
  \end{array}
\right),
\end{equation}
so the averaged motion reduces to the familiar undamped harmonic oscillation in classical mechanics.

We now evaluate the fluctuation behaviors by introducing covariance matrix
\begin{equation}\label{eq-correl-matrix}
{\sigma}(t) \equiv \left(
  \begin{array}{cc}
    \langle v^2\rangle-\langle v\rangle^2 & \omega(\langle vx\rangle-\langle v\rangle \langle x\rangle) \\
    \omega (\langle x v\rangle-\langle x\rangle \langle v\rangle) & \omega^2(\langle x^2\rangle-\langle x\rangle^2) \\
  \end{array}
\right).
\end{equation}
Using Eqs.~(\ref{eq-noisezetamean1D2}), (\ref{eq-zetacorrelate1D2}), and the solution given by Eq.~(\ref{eq-oscilator}), we find that the covariance matrix evolves with time as follows:
\begin{equation}\label{eq-correl-matrixevolu}
\sigma(t)=U\sigma(0) U^\mathrm{T}+\frac{g_{22}}{2\omega} \left(
  \begin{array}{cc}
    \omega t+\frac{\sin 2\omega t}{2} & \frac{1-\cos 2\omega t}{2}\\
    \frac{1-\cos 2\omega t}{2}  & \omega t-\frac{\sin 2\omega t}{2}  \\
  \end{array}
\right)
\end{equation}
The second term on the right-handed side of the above equation is the signature of stochastically broken reciprocity. When $g_{22}=0$, Eq.~(\ref{eq-correl-matrixevolu}) returns to the familiar evolution of the covariance matrix for oscillators in classical mechanics.

\section{PDFs, energy and entropy\label{sec-evolpdf}}
In this section, we will derive three Fokker-Planck equations characterizing the evolutions of PDFs for the COM motion, the two-body relative motion, and their joint evolution, respectively. With these PDFs we then examine how stochastically broken reciprocity gives rise to both energy gain and entropy production in the system.

\subsection{PDF for the COM motion}
Since equations of motion of the COM, (\ref{eq-motCM1DX}) and (\ref{eq-motCM1DV}), are stochastic, we may define a PDF $\rho(X,V,t)$ such that $\rho(X,V,t)\mathrm{d}X\mathrm{d}V$ represents the probability of finding the COM position and velocity at time $t$ within the small rectangle with edges $\mathrm{d}X$ and $\mathrm{d}V$ centered at the phase point $(X,V)^\mathrm{T}$.

Deriving the evolution of the PDF from the stochastic equations of motion is a standard exercise in the textbook of modern statistical physics~\cite{Reichlbook}. Following the method outlined in Ref.~\cite{Reichlbook}, we can straightforwardly derive the Fokker-Planck equation
\begin{equation}\label{eq-FK-COM}
\frac{\partial \rho}{\partial t}=-V\frac{\partial \rho}{\partial X} +\frac{g_{11}}{2}\frac{\partial^2 \rho}{\partial V^2}
\end{equation}
corresponding to Eqs.~(\ref{eq-motCM1DX}) and (\ref{eq-motCM1DV}). This equation describes the evolution of the PDF of the COM.

\subsection{PDF for the two-body relative motion}
The two-body relative motion is governed by equations (\ref{eq-relatposi}) and (\ref{eq-relatvelc}). we may define a PDF $\varrho(x,v,t)$ such that $\varrho(x,v,t)\mathrm{d}x\mathrm{d}v$ represents the probability of finding the relative position and velocity at time $t$ within the small rectangle with edges $\mathrm{d}x$ and $\mathrm{d}v$ centered at the phase point $(x,v)^\mathrm{T}$. Following the method in Ref.~\cite{Reichlbook}, we can easily derive the Fokker-Planck equation
\begin{equation}\label{eq-FK-relative}
\frac{\partial \varrho}{\partial t}=-v\frac{\partial \varrho}{\partial x}+\frac{\phi'}{\mu} \frac{\partial \varrho}{\partial v} +\frac{g_{22}}{2}\frac{\partial^2 \varrho}{\partial v^2},
\end{equation}
corresponding to Eqs.~(\ref{eq-relatposi}) and (\ref{eq-relatvelc}). This equation describes the evolution of the PDF of the two-body relative motion.

\subsection{Joint PDF for the COM motion and the relative motion}
We define a joint PDF $f(X,V,x,v,t)$ such that $f(X,V,x,v,t) \mathrm{d}\Gamma$ represents the probability of finding the COM position and velocity, and the relative position and velocity of B with respect to A, within the small element $\mathrm{d}^4\Gamma\equiv \mathrm{d}X\mathrm{d}V\mathrm{d}x\mathrm{d}v$ centered at the phase point $\bm{\Gamma}\equiv (X,V,x,v)^\mathrm{T}$. Since Eq.~(\ref{eq-g12}) implies $\langle\zeta_1(t)\zeta_2(t)\rangle\neq 0$, the joint PDF $f(X,V,x,v,t)$ usually differs from $\rho(X,V,t)\varrho(x,v,t)$. We may derive the evolution of the joint PDF following the standard stochastic method in Ref.~\cite{Gardinerbook}.

Considering the transformation relations (\ref{eq-zeta1}) and (\ref{eq-zeta2}), we can rewrite Eqs.~(\ref{eq-motCM1DX}), (\ref{eq-motCM1DV}), (\ref{eq-relatposi}) and (\ref{eq-relatvelc}) in a compact form as
\begin{equation}\label{eq-jointmotion}
\mathrm{d}\bm{\Gamma} =\bm{\Lambda} \mathrm{d} t + \Omega \mathrm{d} \mathbf{W}(t)
\end{equation}
where the deterministic phase velocity $\bm{\Lambda}$ and the noise strength matrix $\Omega$ can be explicitly expressed as
\begin{equation}\label{eq-phaseveloc}
\bm{\Lambda}=(V,0,v,-\phi'/\mu)^\mathrm{T}
\end{equation}
and
\begin{equation}\label{eq-matrixomega}
\Omega = \left(
           \begin{array}{cc}
             0 & 0 \\
             \sqrt{g_\mathrm{A}}/M & \sqrt{g_\mathrm{B}}/M \\
             0 & 0 \\
             -\sqrt{g_\mathrm{A}}/m_\mathrm{A} & \sqrt{g_\mathrm{B}}/m_\mathrm{B} \\
           \end{array}
         \right).
\end{equation}
The noise term $\mathrm{d}\mathbf{W}(t)= \left(\xi_\mathrm{A}(t) \mathrm{d}t, \xi_\mathrm{B}(t) \mathrm{d}t\right)^\mathrm{T}$ represents two-variable Wiener process.

According to the method in Ref.~\cite{Gardinerbook}, we can write the corresponding Fokker-Planck equation as
\begin{equation}\label{eq-FKjoint}
\frac{\partial f}{\partial t}=-\nabla_\Gamma\cdot\mathbf{J},
\end{equation}
with the flux
\begin{eqnarray}
\mathbf{J}&=&\bm{\Lambda}f-\frac{1}{2}\Omega \Omega^\mathrm{T}\nabla_\Gamma f\nonumber\\
&=&\left(
     \begin{array}{c}
       Vf \\
       -\frac{g_{11}}{2}\frac{\partial f}{\partial V}-\frac{g_{12}}{2}\frac{\partial f}{\partial v} \\
       vf \\
       -\frac{\phi\prime}{\mu}f -\frac{g_{12}}{2}\frac{\partial f}{\partial V}-\frac{g_{22}}{2}\frac{\partial f}{\partial v} \\
     \end{array}
   \right)\label{eq-flux}
\end{eqnarray}
where $g_{11}$, $g_{12}$ and $g_{33}$ are coefficients shown in Eqs.~(\ref{eq-g11})--(\ref{eq-g22}). Note that the symbol $\nabla_\Gamma$ represents the gradient operator on the phase space $\{\bm{\Gamma}\equiv (X,V,x,v)^\mathrm{T}\}$.

With the above evolution equations of the joint PDF, we can easily verify Eqs.~(\ref{eq-FK-COM}) and (\ref{eq-FK-relative}) using $\rho(X,V,t)=\int f(X,V,x,v,t) \mathrm{d}x\mathrm{d}v$, $\varrho(x,v,t)=\int f(X,V,x,v,t) \mathrm{d}X\mathrm{d}V$, and the Stokes' theorem familiar in multivariable calculus.

\subsection{Energy gain}
Based on the experience of stochastic thermodynamics~\cite{Sekimotobook,Seifert126001,LiTuPRE012127}, we may define the energy as the average mechanical energy in classical mechanics, which reads
\begin{eqnarray}
E&=&\left\langle\frac{M}{2}V^2+\frac{\mu}{2}v^2+\phi(x)\right\rangle\nonumber\\
&=&\int \left[\frac{M}{2}V^2+\frac{\mu}{2}v^2+\phi(x)\right] f \mathrm{d}^4\Gamma \nonumber\\
&=&\frac{M}{2} \int V^2 \rho \mathrm{d}X \mathrm{d}V+\int \left[\frac{\mu}{2}v^2+\phi(x)\right] \varrho \mathrm{d}x \mathrm{d}v \label{eq-energy1}
\end{eqnarray}

Using Eq.~(\ref{eq-FK-COM}) and the Stokes' theorem, and assuming vanishing boundary integrals at infinity, we can obtain
\begin{equation}
\int V^2 \frac{\partial\rho}{\partial t} \mathrm{d}X \mathrm{d}V =g_{11}.
\end{equation}
Similarly, using Eq.~(\ref{eq-FK-relative}) and the Stokes' theorem, and assuming vanishing boundary integrals at infinity, we can obtain
\begin{equation}
\int \left[\frac{\mu}{2}v^2+\phi(x)\right] \frac{\partial\varrho}{\partial t} \mathrm{d}x \mathrm{d}v = \frac{\mu g_{22}}{2}.
\end{equation}
Therefore, we eventually arrive at
\begin{eqnarray}
\frac{\mathrm{d} E}{\mathrm{d}t}&=&\frac{M}{2} \int V^2 \frac{\partial\rho}{\partial t} \mathrm{d}X \mathrm{d}V+\int \left[\frac{\mu}{2}v^2+\phi(x)\right] \frac{\partial\varrho}{\partial t} \mathrm{d}x \mathrm{d}v\nonumber\\
&=&\frac{Mg_{11}+\mu g_{22}}{2}\ge 0.\label{eq-dEdt}
\end{eqnarray}
The equal sign in the above equation holds only for $g_{11}=g_{22}=0$ (corresponding to $g_\mathrm{A}=g_\mathrm{B}=0$). Thus, the stochastically broken reciprocity gives rise to an energy gain in the system. Since no damping terms are included in the minimal model, the stochastic interactions effectively act as an energy input to the system, a behavior analogous to observed phenomena in active matter systems.

\subsection{Entropy production}
Following the concept of stochastic thermodynamics~\cite{Seifert126001}, we define the trajectory entropy as
\begin{equation}\label{eq-trajentropy}
s=-\ln f,
\end{equation}
where the prefactor $k_B$ has been set to unity. The entropy is then regarded as the average of the trajectory entropy, which reads
\begin{equation}\label{eq-entropy}
S=\langle s \rangle=-\int f\ln f \mathrm{d}^4\Gamma.
\end{equation}

The rate of entropy change is
\begin{equation}\label{eq-entropyrate}
\frac{\mathrm{d} S}{\mathrm{d} t}=-\int \frac{\partial f}{\partial t}(\ln f+1) \mathrm{d}^4\Gamma.
\end{equation}
Substituting Eq.~(\ref{eq-FKjoint}) into the above equation, then using the Stokes' theorem, and assuming vanishing boundary integrals at infinity, we obtain
\begin{equation}\label{eq-entropyrate2}
\frac{\mathrm{d} S}{\mathrm{d} t}=\int \mathbf{J}\cdot \nabla_\Gamma (\ln f+1) \mathrm{d}^4\Gamma.
\end{equation}
Substituting the expression (\ref{eq-flux}) for the flux $\mathbf{J}$ into the above equation, we can derive
\begin{equation}\label{eq-entropyprod}
\frac{\mathrm{d} S}{\mathrm{d}t}=\frac{1}{2}\int \left(\sum_{i=1}^2\sum_{j=1}^2g_{ij}s_is_j\right)f \mathrm{d}^4\Gamma.
\end{equation}
with $s_1\equiv{\partial s}/{\partial V}$ and $s_2\equiv{\partial s}/{\partial v}$.
From Eqs.~(\ref{eq-g11})--(\ref{eq-g22}), we derive $g_{11}g_{22}-g_{12}^2={g_\mathrm{A}g_\mathrm{B}}/{m_\mathrm{A}^2m_\mathrm{B}^2} \ge 0$. As a consequence, the integrand in Eq.~(\ref{eq-entropyprod}) takes the form of a positive-definite quantity, leading to the result that ${\mathrm{d} S}/{\mathrm{d}t}\ge 0$. The equality condition, ${\mathrm{d} S}/{\mathrm{d}t}=0$, holds exclusively when $g_\mathrm{A}=g_\mathrm{B}=0$, and this is rigorously proven in Appendix~\ref{sec-appendB}. Therefore, it can be concluded that the stochastic breaking of reciprocity gives rise to the entropy production in the two-body system. From a physical perspective, non-reciprocity disrupts the principle of detailed balance, compelling the system to persist in a sustained non-equilibrium state. Entropy production thus emerges as a natural and inevitable consequence of this non-equilibrium state.

\section{Two-body system immersed in a thermal bath\label{sec-inbaths}}
We further ask what will happen if we place the two-body system mentioned above in a thermal bath at a constant temperature $T$. For simplicity, we set $k_B$ to unity and discuss the overdamped situation.

The equations of motion can be expressed as
\begin{numcases}{}
 \gamma_\mathrm{A}\dot{\mathbf{r}}_\mathrm{A}  =\nabla \phi(r)+\sqrt{g_\mathrm{A}}\bm{\xi}_\mathrm{A}(t)+\sqrt{2\gamma_\mathrm{A} T}\bm{\xi}_\mathrm{TA}(t), \label{eq-motAoverdam}\\
 \gamma_\mathrm{B}\dot{\mathbf{r}}_\mathrm{B} =-\nabla \phi(r)+\sqrt{g_\mathrm{B}}\bm{\xi}_\mathrm{B}(t)+\sqrt{2\gamma_\mathrm{B} T}\bm{\xi}_\mathrm{TB}(t), \label{eq-motBoverdam}
\end{numcases}
where $\gamma_\mathrm{A}$ and $\gamma_\mathrm{B}$ are the damping coefficients of particles A and B, respectively. The terms $\sqrt{2\gamma_\mathrm{A} T}\bm{\xi}_\mathrm{TA}(t)$ and $\sqrt{2\gamma_\mathrm{B} T}\bm{\xi}_\mathrm{TB}(t)$ represent thermal noises due to the bath, which are assumed to be Gaussian white noise.
Additionally, we assume that $\bm{\xi}_\mathrm{A}(t)$, $\bm{\xi}_\mathrm{B}(t)$, $\bm{\xi}_\mathrm{TA}(t)$, and $\bm{\xi}_\mathrm{TB}(t)$ are independent of each other.

It is noted that Eqs.~(\ref{eq-motAoverdam}) and (\ref{eq-motBoverdam}), along with their underdamped counterparts, have been introduced by Kumar~\textit{et al.}~\cite{RamaswamyPRE020102} and Baule~\textit{et al.}~\cite{RamaswamyJSMP11008} in the context of the Brownian inchworm model for self-propulsion. The exact solutions to these equations and comprehensive discussions on this model can be found in Refs.~\cite{RamaswamyPRE020102} and~\cite{RamaswamyJSMP11008}. Herein, we outline only the key results regarding the two-body relative motion.

From Eq.~(\ref{eq-motAoverdam}) and (\ref{eq-motBoverdam}), we can derive the equation of two-body relative motion
\begin{equation}\label{eq-relmotoverdam}
\dot{\mathbf{r}}=-\frac{\nabla \phi}{\nu} +\bm{\chi}(t),
\end{equation}
where the reduced damping coefficient $\nu\equiv {\gamma_\mathrm{A}\gamma_\mathrm{B}}/({\gamma_\mathrm{A}+\gamma_\mathrm{B}})$, and the noise term is given by $\bm{\chi}(t)=({\sqrt{g_\mathrm{B}}}/{\gamma_\mathrm{B}})\bm{\xi}_\mathrm{B}(t)+\sqrt{{2 T}/{\gamma_\mathrm{B}}}\bm{\xi}_\mathrm{TB}(t)-({\sqrt{g_\mathrm{A}}}/{\gamma_\mathrm{A}})\bm{\xi}_\mathrm{A}(t)-\sqrt{{2 T}/{\gamma_\mathrm{A}}}\bm{\xi}_\mathrm{TA}(t)$. It is not hard to verify that
\begin{equation}\label{eq-noisechi-mean}
\left\langle\bm{\chi}(t)\right\rangle=0,
\end{equation}
and
\begin{equation}\label{eq-noisechi-cor}
\left\langle\bm{\chi}(t)\bm{\chi}^\mathrm{T}(t')\right\rangle=2D_e \mathbf{I}\delta(t-t')
\end{equation}
with $D_e\equiv{T}/\nu+{{g_\mathrm{A}}}/{2\gamma_\mathrm{A}^2}+{{g_\mathrm{B}}}/{2\gamma_\mathrm{B}^2}$. Assuming that the Einstein relation~\cite{Reichlbook} still holds, we may define an effective temperature
\begin{equation}\label{eq-effectiveT}
T_{e}\equiv\nu D_e= {T}+\frac{\nu}{2}\left(\frac{{g_\mathrm{A}}}{\gamma_\mathrm{A}^2}+\frac{{g_\mathrm{B}}}{\gamma_\mathrm{B}^2}\right),
\end{equation}
which is clearly larger than the bath temperature $T$ since $\nu$, $g_\mathrm{A}$, $g_\mathrm{B}$ are positive quantities.

Following the method sketched in Ref.~\cite{Reichlbook}, we can readily derive the Smoluchowski equation corresponding to Eqs.~(\ref{eq-relmotoverdam})--(\ref{eq-noisechi-cor}). The PDF $P(\mathbf{r},t)$ for the two-body relative motion satisfies
\begin{equation}\label{eq-FKoverdampt}
\frac{\partial P}{\partial t}=D_e \nabla\cdot \left[\frac{(\nabla \phi) {P}}{T_{e}}+\nabla P\right].
\end{equation}
From the above equation, we observe that a steady state exists. Particularly, the steady-state PDF follows the Boltzmann distribution as
\begin{equation}\label{eq-steadypdf}
P\propto \exp\left\{-\frac{\phi(x)}{T_{e}}\right\}.
\end{equation}

\section{Extremely minimal model\label{sec-mostminmodel}}
In the minimal model mention above, $\bm{\xi}_\mathrm{A}(t)$ and $\bm{\xi}_\mathrm{B}(t)$ are assumed to be independent of each other. We may heuristically consider an extreme situation in which $\bm{\xi}_\mathrm{A}(t)$ and $\bm{\xi}_\mathrm{B}(t)$ are not independent such that $\bm{\zeta}_2(t)$ in Eq.~(\ref{eq-zeta2}) is vanishing. In this sense, the two-body system is referred to as the extremely minimal model.

By setting Eq.~(\ref{eq-zeta2}) to be vanishing, we obtain a necessary condition: $\frac{\sqrt{g_\mathrm{A}}}{m_\mathrm{A}}\bm{\xi}_\mathrm{A}(t)=\frac{\sqrt{g_\mathrm{B}}}{m_\mathrm{B}}\bm{\xi}_\mathrm{B}(t)$, which enlightens us to assume $\sqrt{g_\mathrm{A}}\bm{\xi}_\mathrm{A}(t)=m_\mathrm{A}\sqrt{g}\bm{\xi}(t)$ and $\sqrt{g_\mathrm{B}}\bm{\xi}_\mathrm{B}(t)=m_\mathrm{B}\sqrt{g}\bm{\xi}(t)$ where $g$ is a positive constant quantity. To guarantee this assumption, we need to revisit the main equations in Sec.~\ref{sec-minmodel}. Specifically, equations~(\ref{eq-FBA}) and (\ref{eq-FAB}) are rewritten as
\begin{numcases}{}
 \mathbf{F}_\mathrm{BA} =-\nabla \phi(r)+m_\mathrm{B}\sqrt{g}\bm{\xi}(t),\label{eq-FBAext}\\
 \mathbf{F}_\mathrm{AB} =\nabla \phi(r)+m_\mathrm{A}\sqrt{g}\bm{\xi}(t).\label{eq-FABext}
\end{numcases}
Equations~(\ref{eq-noisemean}) and (\ref{eq-noisecorrelate}) are replaced with
\begin{equation}\label{eq-noisemeanrev}
\left\langle\bm{\xi} (t)\right\rangle=0,
\end{equation}
and
\begin{equation}\label{eq-noisecorrelaterev}
\left\langle\bm{\xi}(t)\bm{\xi}^\mathrm{T}(t')\right\rangle=\delta(t-t')\mathbf{I}.
\end{equation}
The equations governing the COM motion and the relative motion of B with respect to A are revised to
\begin{numcases}{}
\ddot{\mathbf{R}}= \sqrt{g}\bm{\xi}(t), \label{eq-motCMrev}\\
\ddot{\mathbf{r}}= -\frac{\nabla \phi}{\mu} . \label{eq-motrelativerev}
\end{numcases}
Thus, this system is more concise, as the relative motion follows the same rule as in classical mechanics. The COM maintains uniform rectilinear motion on average. The stochastically broken reciprocity only affects the fluctuating motion of the COM. The main conclusions in Secs.\ref{sec-correfunc}A and \ref{sec-evolpdf}A remain unchanged. The other conclusions need to be reexamined in detail.

\section{conclusion}

In the above discussion, we have proposed a minimal model consisting of a two-body system with stochastically broken reciprocity. Guided by Newton's second law, we have derived two stochastic differential equations that govern the COM motion and the two-body relative motion, respectively. Based on these equations of motion, we have obtained the fluctuations of velocity and position for the COM motion and the two-body relative motion, respectively. Additionally, we have derived three Fokker-Planck equations, which respectively characterize the evolution of PDFs for the COM motion, the two-body relative motion, and the joint evolution of these two motions. Using these Fokker-Planck equations, we have analyzed the features of energy gain and entropy production in this two-body system arising from the stochastically broken reciprocity. We have further explored the two-body system in a constant-temperature thermal bath and determined the effective temperature of the system under the overdamped condition. The corresponding Smoluchowski equation, which describes the evolution of the PDF of the two-body relative position, yields the Boltzmann distribution with the effective temperature in the steady state. Finally, we have introduced an extremely minimal model where the relative motion adheres to the laws of classical mechanics, and the effect of stochastically broken reciprocity is solely manifested in the fluctuating motion of the COM.

Before concluding this paper, we discuss three prospective issues for future research:

1) Experimental verification.  A key question is whether the minimal model with stochastically broken reciprocity can be experimentally tested. As we have predicted, the mean square velocity (\ref{eq-corlateV2}) and the mean square displacement (\ref{eq-corlateXX}) contain linear and cubic terms dependent of time, respectively. These distinct scaling laws offer tangible observables for experimental investigation, specifically through tracking the fluctuating motion of the COM. In practical implementation, such an experiment could be designed as follows: observe two small beads in outer space, record the trajectory of their COM, then compute the mean square displacement from the trajectory data and analyze its long-time characteristics. Observation of the cubic scaling would serve as a signature of stochastically broken reciprocity. It should be noted that the effects induced by stochastically broken reciprocity are anticipated to be extremely small in magnitude. Consequently, a critical challenge for experimental physicists will lie in effectively isolating these subtle effects from the confounding influences of other objects or environmental factors.

2) Mechanism of stochastic two-body interaction. We have proved that stochastically broken reciprocity induces an energy gain in two-body systems. However, at the current stage, we are unable to provide a definitive explanation for the energy source, as the microscopic mechanism governing stochastically broken reciprocity remains elusive. If some experimental validations support our model, we could argue that stochastic two-body interaction originates from the vacuum--conceptualized as an active ``aether". In this context, the energy source would be attributed to vacuum fluctuations.

3) Extension to many-body systems. The framework of the present minimal model can be extended to many-body systems, where pairwise interactions are assumed to stochastically violate Newton's third law following the same mechanism as in Eqs.~(\ref{eq-FBA}) and (\ref{eq-FAB}). If no additional constraints are imposed on the entire system, the COM motion is expected to follow a rule similar to Eq.~(\ref{eq-motCM}), meaning the fluctuation behaviors for the COM motion (discussed in Sec. \ref{sec-correfunc}A) would remain valid. However, the motion of particles relative to the COM would be far more complex than in the two-body case, warranting further investigation.

\section*{ACKNOWLEDGMENTS}
The author is grateful for Haijun Zhou for useful discussions, and thanks Luca Cocconi
and Ananyo Maitra for drawing attention to references~\cite{CocconiPRS2023}, \cite{RamaswamyPRE020102} and \cite{RamaswamyJSMP11008}. This work is supported by the National Natural Science Foundation of China (Grant No. 12475032).

\appendix
\section{\label{sec-appendA}Expectation of the Product of a Coordinate and Noise}

Let us consider a simple dimensionless one-dimensional stochastic differential equation given by
\begin{equation}\dot{x}=-[1+\xi(t)]x\label{eq-apa1}\end{equation}
where $\xi(t)$ is a Gaussian white noise. We aim to address the subtle issue of whether $\langle\xi(t){x}(t)\rangle$ vanishes or not?

To tackle this question, we first solve Eq.~(\ref{eq-apa1}) to obtain the expression for $x(t)$. Integrating the differential equation yields
\begin{equation}{x}=x_0\mathrm{e}^{-\int_0^t[1+\xi(\tau)]d\tau}=x_0\mathrm{e}^{-t}\mathrm{e}^{-\int_0^t\xi(\tau)d\tau}\label{eq-apa2}\end{equation}
where $x_0$ denotes the initial condition at $t=0$.

Next, we multiply both sides of Eq.~(\ref{eq-apa2}) by $\xi(t)$ and take the expectation value:
\begin{equation}\langle\xi(t){x}(t)\rangle=\langle x_0\rangle \mathrm{e}^{-t} \langle \xi(t)\mathrm{e}^{-\int_0^t\xi(\tau)d\tau}\rangle\label{eq-apa4}.\end{equation}
In this step, we have exploited the fact that the initial condition $x_0$ and the noise term $\xi(t)$ are independent of each other.

To evaluate the expectation $\langle \xi(t)\mathrm{e}^{-\int_0^t\xi(\tau)d\tau}\rangle$, we expand the exponential term in a power series:
\begin{eqnarray}
\langle \xi(t)\mathrm{e}^{-\int_0^t\xi(\tau)d\tau}\rangle=\sum_{n=0}^{\infty}\langle \xi(t)(-1)^n\frac{[\int_0^t\xi(\tau)d\tau]^n}{n!}\rangle \nonumber\\
= -\sum_{n=0}^{\infty}\frac{1}{(2n+1)!} \langle\xi(t) [\int_0^t\xi(\tau)d\tau]^{2n+1} \rangle\label{eq-apa5}
\end{eqnarray}
Here, we note that only the odd-powered terms contribute to the sum, as the even-powered terms vanish due to the symmetry of the Gaussian white noise: the odd moments of $\xi(t)$ are zero.

Since the even moments of the Gaussian white noise are non-negative, $\langle\xi(t) [\int_0^t\xi(\tau)d\tau]^{2n+1} \rangle$ is non-negative. To illustrate this point, we evaluate the first term in the series.
Using the fact that the correlation function of the Gaussian white noise is given by $\langle\xi(t)\xi(\tau)\rangle = \delta(t-\tau)$, where $\delta(t-\tau)$ is the Dirac delta function, we have
\begin{eqnarray}
\langle\xi(t) \int_0^t\xi(\tau)d\tau \rangle=\int_0^t\langle\xi(t)\xi(\tau) \rangle d\tau\nonumber\\
=\int_0^t \delta(t-\tau) d\tau =\frac{1}{2}\label{eq-apa6}
\end{eqnarray}
The last equality can be understood by considering the even parity of the Dirac delta function $\delta(t-\tau)$ about the point $\tau=t$. In addition, the Dirac delta function $\delta(t-\tau)$ is non-zero only for a very narrow domain near $\tau=t$, and its total integral over the entire real line is 1. Thus, we have $\int_0^t \delta(t-\tau) d\tau =\int_{-\infty}^t \delta(t-\tau) d\tau =\int_t^\infty \delta(t-\tau) d\tau=1/2$ for $t>0$.

Similarly, we can verify that all terms in Eq.~(\ref{eq-apa5}) are non-negative. Thus we conclude that
\begin{equation}\langle\xi(t){x}(t)\rangle=\langle x_0\rangle e^{-t} \langle \xi(t)e^{-\int_0^t\xi(\tau)d\tau}\rangle>0\label{eq-apa7}\end{equation}
This result highlights a subtle yet important aspect of stochastic systems with multiplicative noise, where the outcome of an stochastic integral depends on the choice of the integration scheme, such as Ito's or Stratonovich's conventions~\cite{Gardinerbook}.

\section{\label{sec-appendB}Entropy production when $g_\mathrm{A}=0$ but $g_\mathrm{B}>0$}

We examine whether entropy production can vanish in the case where $g_A=0$ while $g_\mathrm{B}>0$.

When $g_A=0$, Eqs.~(\ref{eq-g11})--(\ref{eq-g22}) yield $g_{11}=g_\mathrm{B}/M^2$, $g_{12}=g_{21}=g_\mathrm{B}/Mm_\mathrm{B}$, and $g_{22}=g_\mathrm{B}/m_\mathrm{B}^2$. For the entropy production rate in Eq.~(\ref{eq-entropyrate2}) to vanish, the necessary condition is $s_1/M+s_2/m_\mathrm{B}=0$, which is equivalent to:
\begin{equation}\label{eq-vanentrpy1}
\frac{1}{M}\frac{\partial f}{\partial V}+\frac{1}{m_\mathrm{B}}\frac{\partial f}{\partial v}=0.
\end{equation}

Substituting this condition into Eq.~(\ref{eq-flux}), the corresponding Fokker-Planck equation (\ref{eq-FKjoint}) reduces to:
\begin{equation}\label{eq-eq-FKjointvanga}
\frac{\partial f}{\partial t}+V\frac{\partial f}{\partial X}+v\frac{\partial f}{\partial x}+(-\frac{\phi^\prime}{\mu})\frac{\partial f}{\partial v}=0.
\end{equation}

Under the transformation of variables $V=(m_\mathrm{A}v_\mathrm{A}+m_\mathrm{B}v_\mathrm{B})/M$, $v=v_\mathrm{B}-v_\mathrm{A}$, $X=(m_\mathrm{A}x_\mathrm{A}+m_\mathrm{B}x_\mathrm{B})/M$, and $x=x_\mathrm{B}-x_\mathrm{A}$, the two-particle distribution function is expressed as $\tilde{f}(x_\mathrm{A},v_\mathrm{A},x_\mathrm{B},v_\mathrm{B},t)=f((m_\mathrm{A}x_\mathrm{A}+m_\mathrm{B}x_\mathrm{B})/M,(m_\mathrm{A}v_\mathrm{A}+m_\mathrm{B}v_\mathrm{B})/M,x_\mathrm{B}-x_\mathrm{A},v_\mathrm{B}-v_\mathrm{A},t)$. Eq.~(\ref{eq-vanentrpy1}) is then transformed into
\begin{equation}\label{eq-vanentrpy2}
	\frac{\partial \tilde{f}}{\partial v_\mathrm{B}}=0,
\end{equation}
implying that $\tilde{f}$ is independent of $v_\mathrm{B}$, i.e., $\tilde{f}=\tilde{f}(x_\mathrm{A},v_\mathrm{A},x_\mathrm{B},t)$. With this, Eq.~(\ref{eq-eq-FKjointvanga}) simplifies to:
\begin{equation}\label{eq-eq-FKjointvanga2}
	\frac{\partial \tilde{f}}{\partial t}+v_\mathrm{A}\frac{\partial \tilde{f}}{\partial x_\mathrm{A}}+\frac{\phi^\prime}{m_\mathrm{A}}\frac{\partial \tilde{f}}{\partial v_\mathrm{A}}+v_\mathrm{B}\frac{\partial \tilde{f}}{\partial x_\mathrm{B}}=0.
\end{equation}
The last term in this equation explicitly involves $v_\mathrm{B}$. However, given that $\tilde{f}$ is independent of $v_\mathrm{B}$, the equation can only be satisfied if ${\partial \tilde{f}}/{\partial x_\mathrm{B}}=0$. This implies $\tilde{f}=\tilde{f}(x_\mathrm{A},v_\mathrm{A},t)$, a function dependent solely on the variables of particle A. This conclusion contradicts the fundamental assumption underlying a two-body system, which inherently requires dependence on both particles. Consequently, the necessary condition given by Eq.~(\ref{eq-vanentrpy1}) cannot hold, making $\mathrm{d}S/\mathrm{d}t=0$ impossible. Thus, the only possibility remaining is $\mathrm{d}S/\mathrm{d}t>0$.


\end{document}